\documentclass[twocolumn,reprint,
 superscriptaddress, amsmath,amssymb, aps, prl]{revtex4-1}
\usepackage{graphicx}
\usepackage{amsmath}
\usepackage{color}
\usepackage{dcolumn}
\usepackage{bm}
\usepackage[
   bookmarks=true,
   colorlinks=true,
   hyperindex=true,
   citecolor=blue,
   linkcolor=blue,
   urlcolor=blue
]{hyperref}

\begin{document}
\title{Efficient preparation of 2D defect-free atom arrays with near-fewest sorting-atom moves}
\author{Cheng Sheng}
\affiliation{State Key Laboratory of Magnetic Resonance and Atomic and Molecular Physics, Wuhan
Institute of Physics and Mathematics, APM, Chinese Academy of Sciences , Wuhan 430071, China}
\author{Jiayi Hou}
\affiliation{State Key Laboratory of Magnetic Resonance and Atomic and Molecular Physics, Wuhan
Institute of Physics and Mathematics, APM, Chinese Academy of Sciences , Wuhan 430071, China}
\affiliation{School of Physical Sciences, University of Chinese Academy of Sciences, Beijing 100049, China}
\author{Xiaodong He}
\email{hexd@wipm.ac.cn}
\affiliation{State Key Laboratory of Magnetic Resonance and Atomic and Molecular Physics, Wuhan
Institute of Physics and Mathematics, APM, Chinese Academy of Sciences , Wuhan 430071, China}
\author{Peng Xu}
\affiliation{State Key Laboratory of Magnetic Resonance and Atomic and Molecular Physics, Wuhan
Institute of Physics and Mathematics, APM, Chinese Academy of Sciences , Wuhan 430071, China}
\author{Kunpeng Wang}
\affiliation{State Key Laboratory of Magnetic Resonance and Atomic and Molecular Physics, Wuhan
Institute of Physics and Mathematics, APM, Chinese Academy of Sciences , Wuhan 430071, China}
\author{Jun Zhuang}
\affiliation{State Key Laboratory of Magnetic Resonance and Atomic and Molecular Physics, Wuhan
Institute of Physics and Mathematics, APM, Chinese Academy of Sciences , Wuhan 430071, China}
\affiliation{School of Physical Sciences, University of Chinese Academy of Sciences, Beijing 100049, China}
\author{Xiao Li}
\affiliation{State Key Laboratory of Magnetic Resonance and Atomic and Molecular Physics, Wuhan
Institute of Physics and Mathematics, APM, Chinese Academy of Sciences , Wuhan 430071, China}
\author{Min Liu}
\affiliation{State Key Laboratory of Magnetic Resonance and Atomic and Molecular Physics, Wuhan
Institute of Physics and Mathematics, APM, Chinese Academy of Sciences , Wuhan 430071, China}
\author{Jin Wang}
\affiliation{State Key Laboratory of Magnetic Resonance and Atomic and Molecular Physics, Wuhan
Institute of Physics and Mathematics, APM, Chinese Academy of Sciences , Wuhan 430071, China}
\author{Mingsheng Zhan}
\email{mszhan@wipm.ac.cn}
\affiliation{State Key Laboratory of Magnetic Resonance and Atomic and Molecular Physics, Wuhan
Institute of Physics and Mathematics, APM, Chinese Academy of Sciences , Wuhan 430071, China}
\date{\today}

\begin{abstract}
Sorting atoms stochastically loaded in optical tweezer arrays via an auxiliary mobile tweezer is an efficient approach to preparing intermediate-scale defect-free atom arrays in arbitrary geometries. However, high filling fraction of atom-by-atom assemblers is impeded by redundant sorting moves with imperfect atom transport, especially for scaling the system size to larger atom numbers. Here, we propose a new sorting algorithm (heuristic cluster algorithm, HCA) which provides near-fewest moves in our tailored atom assembler scheme and experimentally demonstrate a $5\times6$ defect-free atom array with 98.4(7)$\%$ filling fraction for one rearrangement cycle. The feature of HCA that the number of moves $N_{m}\approx N$ ($N$ is the number of defect sites to be filled) makes the filling fraction uniform as the size of atom assembler enlarged. Our method is essential to scale hundreds of assembled atoms for bottom-up quantum computation, quantum simulation and precision measurement.
\end{abstract}

\maketitle


\maketitle

\section{Introduction}

Bottom-up builded single atom arrays have rapidly developed into a versatile platform for quantum many-body simulation~\cite{Bernien2017,Keesling2019,deLeseleuc2018,Browaeys2020}, quantum metrology~\cite{Norcia2019,Madjarov2019} and quantum computation~\cite{Xia2015,Wang2017,Sheng2018,Levine2019,Graham2019,Madjarov2020}. This is mainly owing to several crucial advantages, besides configurable defect-free atom arrays via atom rearrangement~\cite{Kim2016,Endres2016,Barredo2016,Lee2017,Brown2019,Mello2019}, long coherence time of quantum bits (qubits) well-isolated from environment~\cite{Yang2016,Guo2020,Li2019,Norcia2019} and controllable long-range interactions using Rydberg atoms~\cite{Urban2009,Saffman2010,Wilk2010,Zeng2017}. Since a great amount of quantum science and technology based on assembled-atom platforms will strongly benefit from scaling the system size to larger atom numbers such as building an fault-tolerant information processor with sufficient logic qubits to solve classical intractable tasks~\cite{Weiss2017,Saffman2016}. Determined preparing hundreds of atoms with full control over spatial geometries is important but still a challenge.

Recent years, the atom-by-atom assemblers have been scaled from one dimension to three dimension~\cite{Barredo2018,Kumar2018}, meanwhile the array size has been increased to about two hundred atoms~\cite{Wang2020}. An effective and conventional approach to achieving the defect-free atom arrays in arbitrary geometries is to sort randomly distributed atoms to target sites via an auxiliary mobile tweezer (MT)~\cite{Barredo2016,Barredo2016,Mello2019}. Generally, a rearrangement for one atom contains extracting the atom out by MT with deeper trap depth, moving the atom in MT along a certain path and releasing it into the target site. Therefore, the filling fractions of the final atom arrays closely relate to atom transfer efficiency. Because the failure of extracting atoms out by a MT makes the atoms stay \emph{in situ} and may cause collisional loss in next transfer process. Moreover, the atom loss in the MT and the imperfect of releasing atoms into target sites lead to defects in final atom arrays. To increase the filling fraction, an optimal sorting path with number of moves $N_{m}$ as few as possible can minimize the influence of the imperfection during the atom rearrangement. Therefore, finding an atom sorting algorithm well suited for large scale defect-free atom assemblers is essential.

A sorting algorithm called heuristic path-finding algorithm (HPFA) is proposed to prepare a $7\times7$ atom array with a filling fraction of 96$\%$ for a $5\times5$ atom array~\cite{Barredo2016}. A similar heuristic shortest-move algorithm has been applied to rearrange 111 atoms and the cumulative success rate of a $4\times4$ atom array is less than $20\%$ for one rearrangement cycle mainly limited by a 75$\%$ transfer efficiency (might be induced by the short lifetime in the MT)~\cite{Mello2019}. However, these sorting algorithms are not optimal for large scaled atom arrays due to a number of redundant moves involved. Although finding the sorting path with fewest is a difficult computational task (likely to be a NP-complete problem). Here, we propose a heuristic cluster algorithm (HCA) which significantly reduces the $N_{m}$ for sorting atoms and demonstrate it experimentally by preparing a $5\times6$ atom array with 98.4(7)$\%$ filling fraction (after correcting the 4$\%$ atom loss due to the background gas collisions and the loss during probing atom process) for one rearrangement cycle.


\begin{figure*}[htbp]
\centering
\includegraphics[width=18cm]{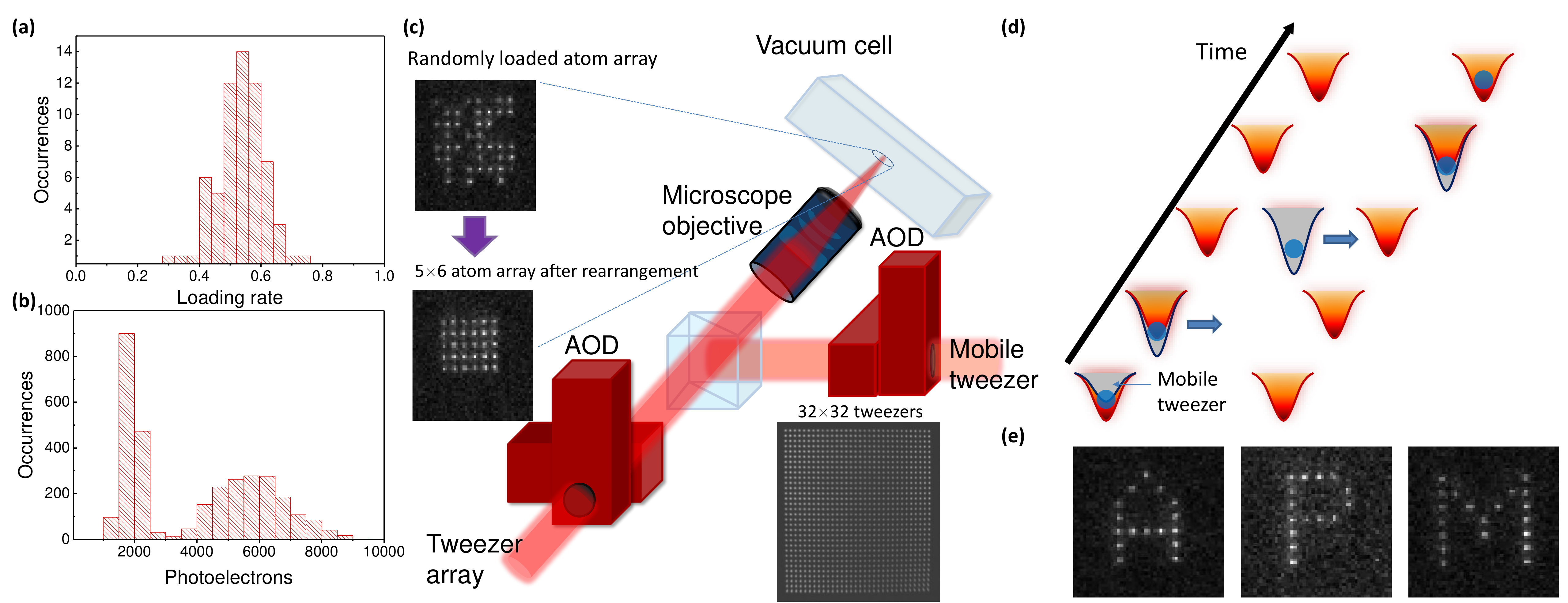}
\caption{(color online). Images of 2D defect-free atom arrays. (a) Loading rates histogram of 64 tweezers. The average loading rate of 64 tweezers is 53$\%$ measured by 200 times repeating loading process. (b) Fluorescence count distribution for 64 tweezers during 80 ms of exposure after 200 times loading process. The region of interest for each tweezer is $2\times2$ pixels. (c) Schematics of the experimental setup. (d) Determined preparing a filled tweezer in target sites via transporting a single atom (blue disk) from another filled tweezer with the MT. This process represents one move. (e) Defect-free atom arrays in the specific patterns of capital English letters via atom rearrangement. The unused extra atoms are discard by moving them to positions far away from the field of view. The distance between neighboring atoms is 5 $\mu$m.}
\label{fig:fig1}
\end{figure*}


We begin by introducing an $8\times8$ atom array setup. Then, we improve the atom lifetime in the MT and prove that no obvious atom loss is observed when a atom is transferred across optical tweezer array in a relatively long move distance. In this case, a sorting algorithm provides fewest moves is preferred. Next, three sorting algorithms are briefly described and compared. Finally, we experimentally measure the filling fraction of atom arrays utilizing these algorithms with two different transfer efficiency and give simulating results for large scaled atom arrays.

\section{Experimental preparation of atom assemblers}

The $8\times8$ optical tweezer array is generated by an 808 nm laser beam deflected in two orthogonal directions by a dual axis acousto-optic deflector (AOD). The dual axis AOD are respectively driven by two radio-frequency (RF) signals with 8 tones (the central frequency is 96 MHz and the frequency gap between the neighboring tones is 2 MHz) which is produced by an arbitrary waveform generator (AWG, Keysight M3202A). The optical homogeneity of the trap array is less than 10$\%$ after optimization. The 830 nm MT is also created by an AOD with the RF signal sent from a voltage controlled oscillator (VCO, Mini-circuits ZOS-150+) with the frequency tuned by a direct current (DC) voltage. The Gaussian waist of the each tweezer in the array and MT are both about 1.0 $\mu$m. Our experimental setup is shown schematically in Fig.~\ref{fig:fig1}(c).

Single $^{87}$Rb atoms are stochastically loaded from a magneto-optical trap into optical tweezer arrays with trap depth of $U_{0}/k_{B}\approx$0.8 mK for each tweezer in 950 ms, where $k_{B}$ is the Boltzmann constant. The photoelectron count threshold and the regions of interest for every tweezer are periodly calibrated by a computer program via automatically analyzing a series of atom array images. Images of atom fluorescence are taken by an electron multiplying charge coupled device (EMCCD) camera (Princeton Instruments ProEM+:1024B). Fig.~\ref{fig:fig1}(a) and Fig.~\ref{fig:fig1}(b) respectively illustrate the loading rates and the average threshold of 64 sites. Next, atoms in reservoir tweezers are transferred to the target sites using a MT in a real time sorting path calculated by sorting algorithm. The sorting path is composed of three waveforms to control the intensity and X and Y positions of the MT respectively. For the rearrangement process as depicted in Fig.~\ref{fig:fig1}(d) we spend 1 ms adiabatically increasing the trap depth of the MT to 2.3 mK after the MT overlaps a filled reservoir tweezer. Then, the MT extracts the atom out and moves at the speed of 1 ms/grid (one grid represents the distance between two neighboring sites) across void tweezers to the target site. After the atom in the MT released into the target site in 1 ms, the next rearrangement cycles for other target sites are carried out. Finally, we acquire a new image to reveal the new positions of the atoms in the array as shown in Fig.~\ref{fig:fig1}(e). The total rearrangement time is less than 80 ms for a $5\times6$ atom assembler.

\section{Atom lifetimes in a mobile tweezer}

In order to optimize the filling fraction of defect-free atom arrays, we firstly improve the atom lifetime in a MT because the atom loss directly leads to the defects in the final atom arrays. As described above, the position of the MT is controlled by the frequency of RF signals which are sent into the AOD. We experimentally find that the atom lifetime is affected by the beam-pointing fluctuations of the MT~\cite{Savard1997}. The beam-pointing noise observed by the sideband of the RF signal frequency spectrum causes a heating effect when the noise frequency is about 90 kHz (equal to the atom oscillation frequency in the harmonic trap). Since the heating rate is constant, the survival probability of atom in MT as a function of time can be expressed as~\cite{Tuchendler2008}

\begin{figure}[t]
\centering
\includegraphics[width=8.6cm]{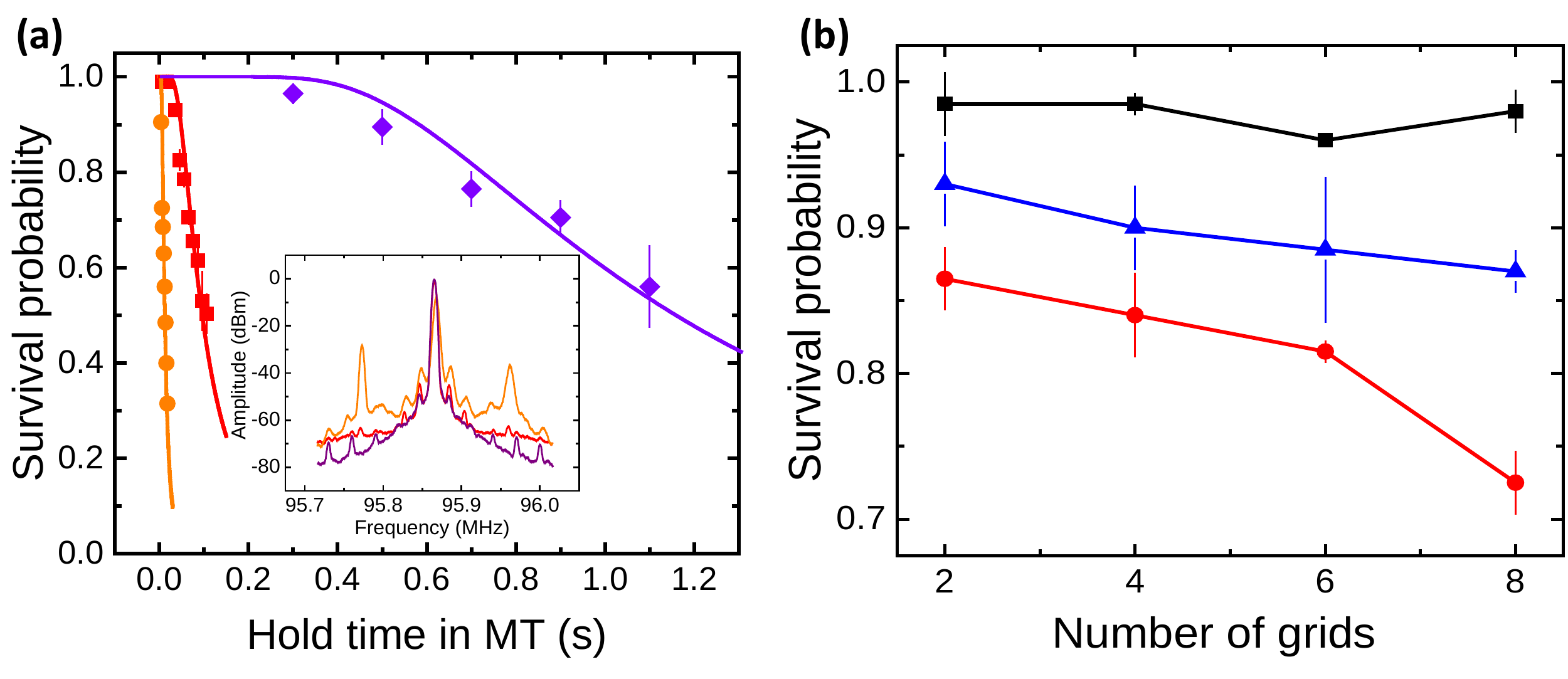}
\caption{(color online). Atom lifetimes in a MT. (a) Atom survival probability is in a static MT. The orange circles, red squares and purple diamonds are respectively denote the lifetime with RF driver generated by three approaches. The first manner is that the RF signals come from a signal generator (Agilent E4433B, the deviation of frequency modulation is 5 MHz/V) with frequency modulated by a DC voltage sent from an analog output (National Instruments PXI-6733). In the second manner, a purer DC voltage is generated by an AWG (Keysight 33522A) instead of the analog output. For the third one, VCO produces the RF signals tuned by DC voltage output from the AWG. The solid curves are fitted to Eq.~(\ref{eq1}) and heating rates are 23 mK/s (orange circles), 2.9 mK/s (red squares), 0.24 mK/s (purple diamonds). The inset illustrates the frequency spectrum of the RF signals. The sidebands are induced by the noise of the DC voltage to control the RF frequency in the mode of frequency modulation. (b) Atom survival probability in a MT after transferred across several void tweezers in the array. The black squares, blue triangles, red circles respectively represent the results with the MT depth of 2.27 mK, 0.82 mK, 0.77 mK and the trap depth of void tweezers is 0.8 mK.}
\label{fig:fig2}
\end{figure}

\begin{equation}\label{eq1}
P_{s}(t)=1-[1+{\nu (t)}+\frac{1}{2}\nu^{2}(t)]e^{-\nu (t)},
\end{equation}
where ${\nu (t)}=U_{0}/[{k_{B}(T_{0}+rt)]}$, and $t$ is the atom hold time in MT, $T_{0}=20$ $\mu$K represents the initial atom temperature and $r$ denotes the heating rate. We measure the atom lifetime in a static MT with RF signal of different frequency spectrum and conclude that the lifetime depends on the amplitude of the sidebands (about 90 kHz detuning from the carrier) as depicted in Fig.~\ref{fig:fig2}(a). Moreover, the lifetime is further increased to 7 s when we apply continuous laser cooling (turning on repump light resonant with $|5s_{1/2}, f=1\rangle\rightarrow|5p_{1/2}, f=1\rangle$ and cooling light red detuned by 24 MHz-54 MHz from $|5s_{1/2}, f=2\rangle\rightarrow|5p_{3/2}, f=3\rangle$ simultaneously). Additionally, we should note that another heating effect due to the light interference between neighboring tweezers in the array reduces atom lifetime to 2.1 s, but this heating effect can also be suppressed by the continuous laser cooling. The cooling process however will destroy the internal state if the atoms are encoded into qubits, thus our work that atom lifetime is improved in a MT without cooling is still necessary.

Furthermore, the atom survival probability in the MT is measured after the MT moves across several void tweezers as shown in Fig.~\ref{fig:fig2}(b). We observe that it decreases as the number of void tweezers is increased when the MT depth is less than or approximately equal to the depth of each void tweezer. However, atoms are preserved well when the MT depth is sufficiently deep. Therefore, we tend to choose a sorting algorithm with fewer $N_{m}$ in tailored our atom assembler scheme and the move distance is not taken into the main consideration.


\section{Sorting-atom algorithms}

\begin{figure}[t]
\centering
\includegraphics[width=8.6cm]{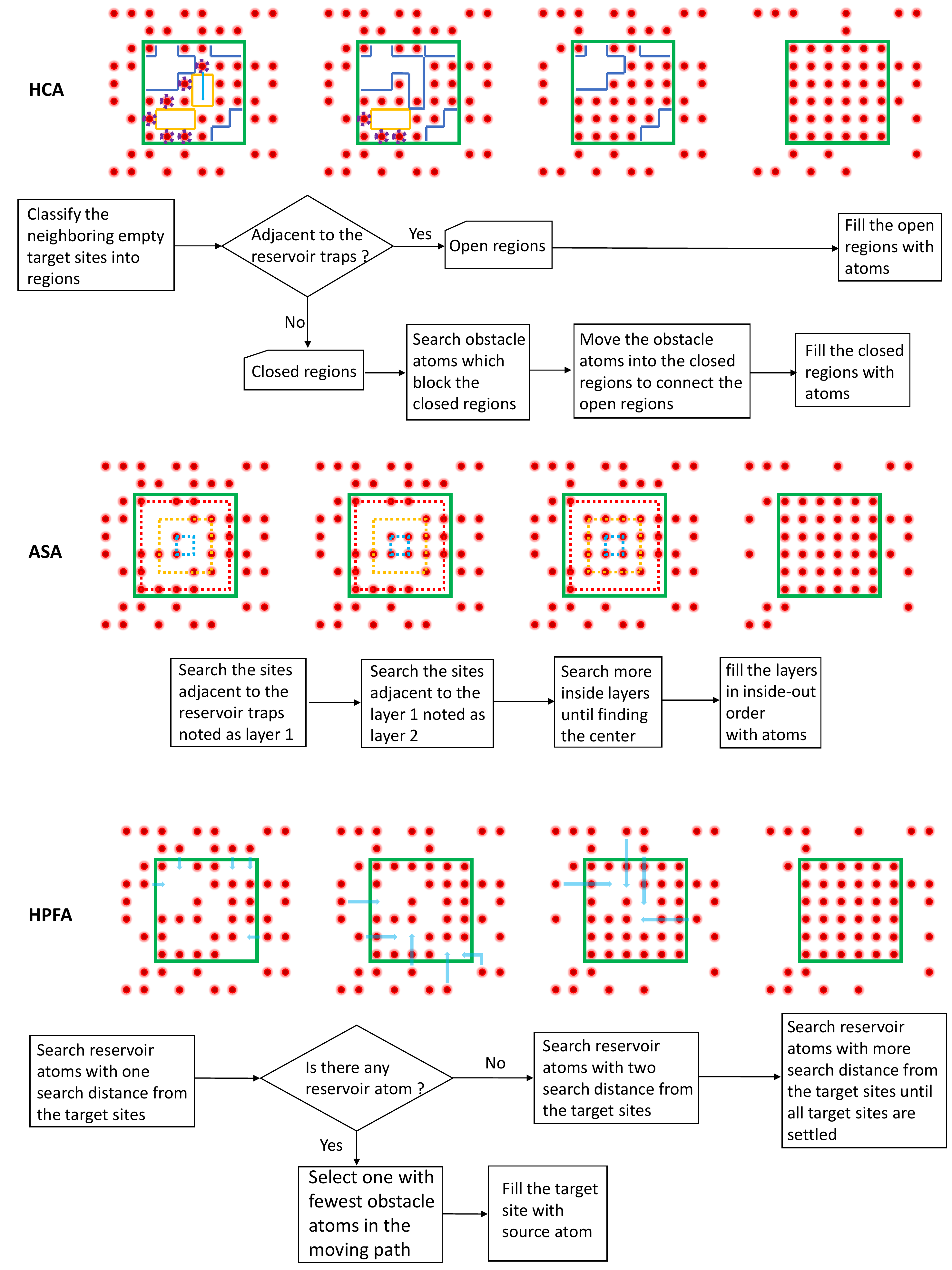}
\caption{(color online). Schematic diagrams and flowcharts of HCA, ASA and HPFA. The sorting process is from left to right.}
\label{fig:fig3}
\end{figure}

Now, we introduce the sorting atom process of HCA, ASA (A* searching algorithm) and HPHA respectively as sketched in Fig.~\ref{fig:fig3}. HCA, aiming at costing the $N_{m}$ as few as possible are proposed to find smart sorting path to fill empty target sites with source atoms in reservoir traps. We call the all reservoir tweezers as the outside-region and all target sites as the inside-region. Three blue sections adjacent to the outside-region and two orange sections isolated from the outside-region are defined as open-regions and closed-regions respectively. The number of defect target sites $N={N}_{1}+{N}_{2}$, where ${N}_{1}$ and ${N}_{2}$ respectively represent the number of filled target sites in open-regions and closed-regions. The total number of moves ${N}_{m}={N}_{o}+{N}_{c}$, where ${N}_{o}$ and ${N}_{c}$ respectively represent the number of moves to fill the open-regions and closed-regions. ${N}_{o}={N}_{1}$, because the open-regions can be directly filled with source atoms in outside-region and ${N}_{o}$ is the fewest number to fill target sites. ${N}_{c}>{N}_{2}$, since we need carry out extra moves to transport the obstacle atoms which block the move path. The core ideal of HCA is to convert the close-regions to open-regions. Thus the first step of HCA is to identify the open-regions and closed-regions. Then, obstacle atoms (in dash circle) that block the connection between the closed-regions and the open-regions are moved to the closed-regions to open them. The obstacle atoms are searched around the open regions site by site (sometime, more than one atom should be moved to open up the close regions). After that, all regions are open and filled with the source atoms in order (the site with the longest distance from the outside-region filled first).


ASA is another sorting-atom algorithm developed by us that fills the empty target sites from the center to the edge with two main steps. The first step is to find the center location of the atom array and classify the layers (section in the blue, orange and red dash line square) in inside-out order. The second is to fill the trap, one layer after the other. Tweezers of target sites adjacent to the reservoir tweezers are defined as the first layer. The second layer contains the sites in inside-region adjacent to the first layer. The classifying layer process continues until no new layer can be found. The last one is the center of the atom array. Usually, since filling from the center significantly reduce the sorting moves, this is the reason why we classify the layers. In the second step, ASA fills the layer one by one in the reverse order of layers. The search source atom process is similar to A* searching algorithm (finding the nearest atom and moving it bypassing the obstacles to the target site in the shortest path).

Compared with the HPFA in Ref.~\cite{Barredo2016}, our HPFA does a certain amount of optimization. The upgraded HPFA does not calculate all the distance between each atom and each target tweezer. Instead, we define a parameter named search distance which is set to be one at the beginning. We check all the empty tweezers whether there are atoms in the reservoir tweezers satisfied with the search distance. If the choices are multiple, we will select the one which has the least obstacle atoms in the sorting path to achieve as few moves as possible. If there is an obstacle atom blocking the sorting path, we move this atom to the target site and replace the obstacle by a source atom. The search distance is continuously increasing until all the target tweezers are settled.

\begin{figure}[t]
\centering
\includegraphics[width=8.6cm]{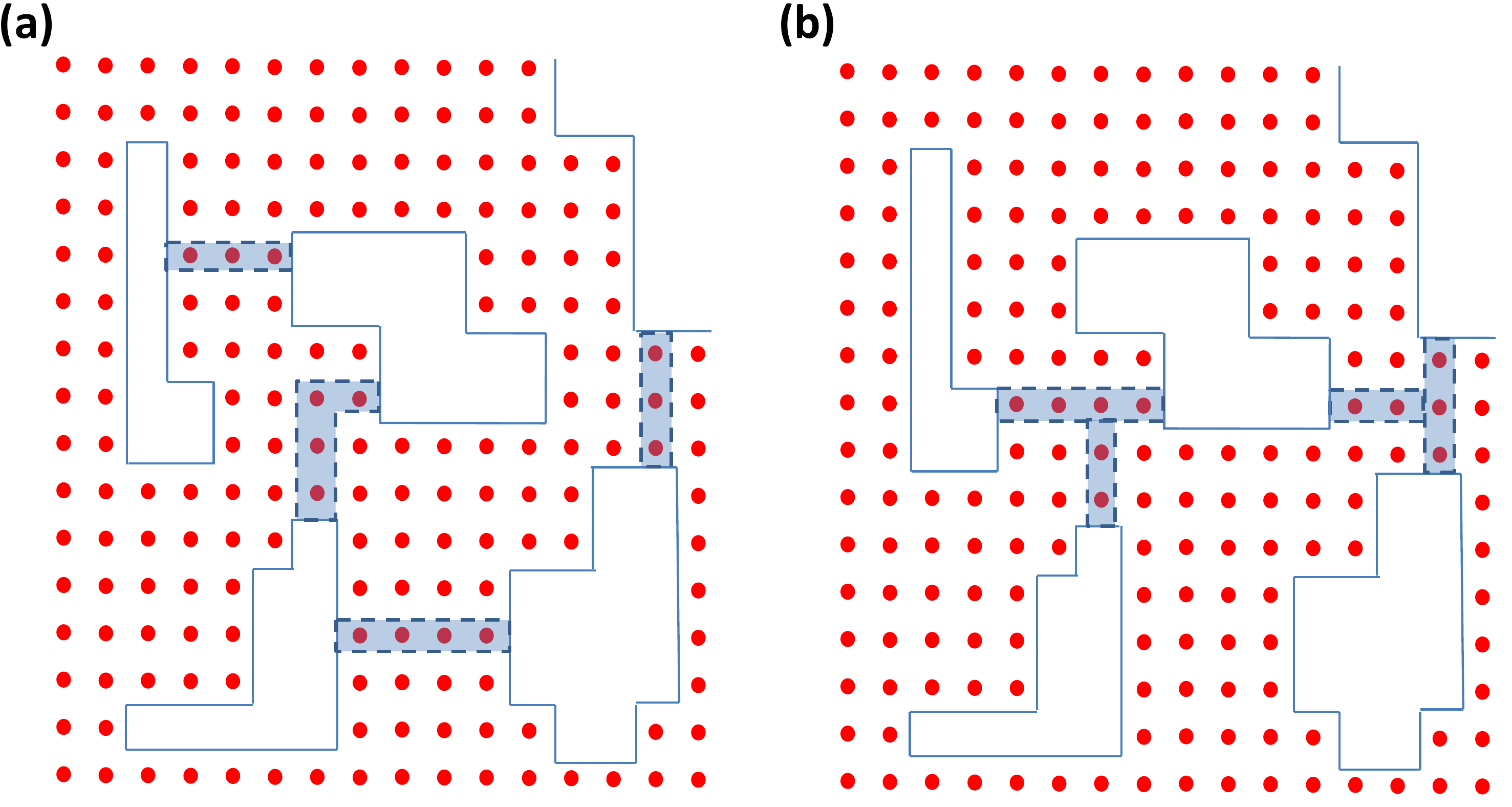}
\caption{(color online). Finding the shortest route to connect all regions. (a) The route only composed of region-region connections with 14 obstacle atoms. (b) The route composed of region-region and region-junction connections with 11 obstacle atoms.}
\label{fig:fig4}
\end{figure}

To further explore the fewest $N_{m}$, we need to find how to connect the all closed-regions and open-regions with shortest route as illustrated in Fig.~\ref{fig:fig4}(a). Because the shortest connecting route means the fewest obstacle atoms should be transported to the closed regions. This task is similar to the traveling salesman problem (TSP) which has been tried to solved by various heuristic algorithms. A region in atom array can be simplified as a site in TSP, and the goal of searching the shortest traveling distance also represents finding the shortest connecting route to open the closed-regions. In fact, the connecting route shown as Fig.~\ref{fig:fig4}(b) is shorter than the route in Fig.~\ref{fig:fig4}(a). Although this indicates finding fewest $N_{m}$ is a more complicated mathematic problem than TSP, some heuristic algorithms to solve TSP applied to upgrade HCA is still a promising way to further minimize $N_{m}$, especially for thousands assembled atoms with high initial loading rate.

\section{Number of moves and filling fractions}


\begin{figure*}[htbp]
\centering
\includegraphics[width=18cm]{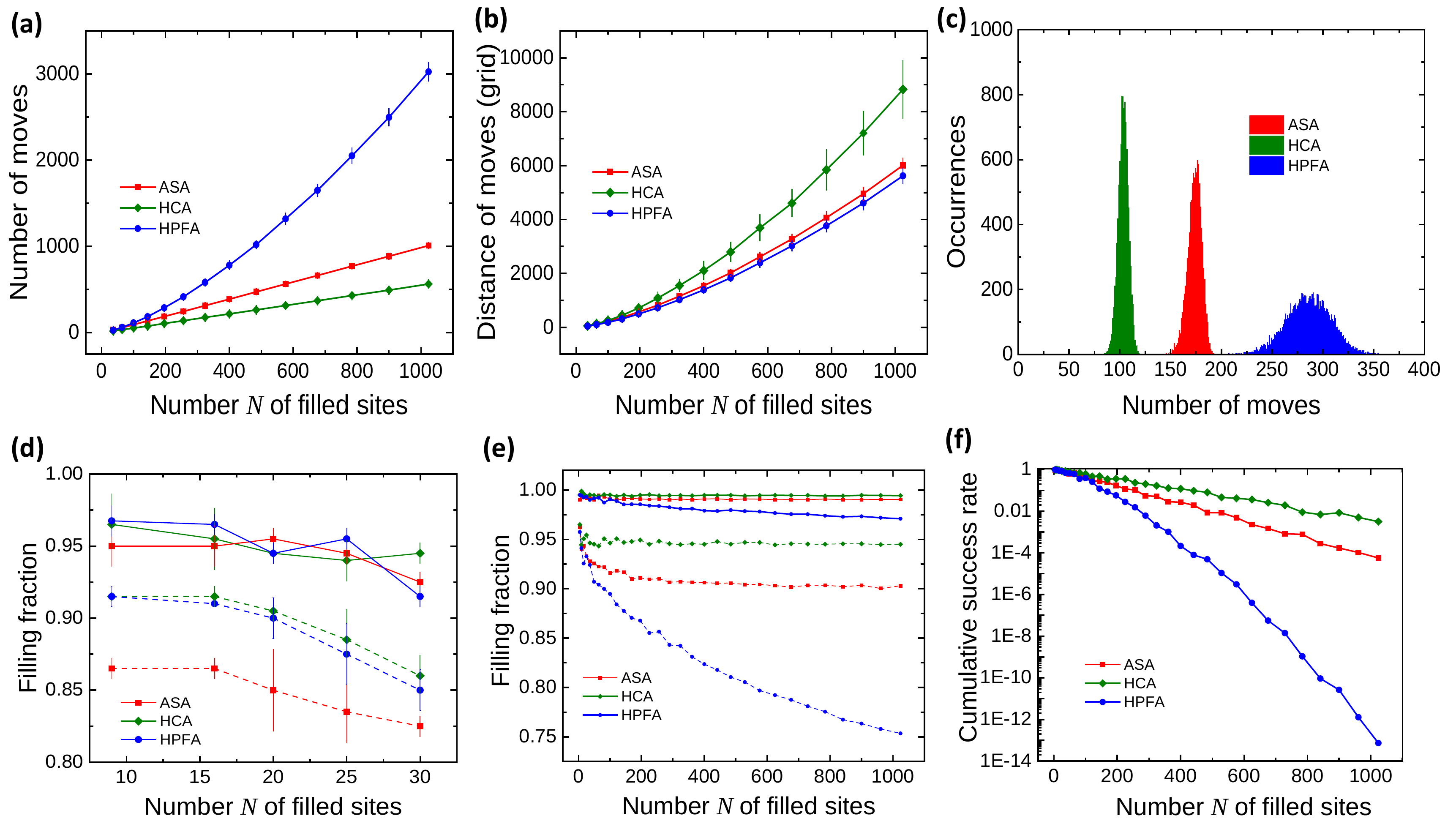}
\caption{(color online). Number of moves of HCA, ASA and HPFA and filling fractions of the defect-free atom array as the number of filled sites. (a) Simulating results of the transfer moves as a function of filled site numbers. (b) Simulating results of the transfer distances as a function of filled site numbers. The error bars are caused by the randomly distributed initial atom arrays with 50$\%$ loading rate. The shape of the target sites are all squares. (c) Histogram of the number of moves to prepare a $14\times14$ square array for 10000 initial random loading instances. (d) Experimental filling fraction of atom array after atom rearranged by ASA, HCA and HPFA. The solid line and the dash line respectively represent the experimental results in the condition of 97.7$\%$ and 88.0$\%$ transfer efficiency (the average values for 64 sites). The lower transfer efficiency is obtained by the position displacement of MT and tweezer array. (e) Simulating filling fraction of large scaled atom array using ASA, HCA and HPFA. The solid line and the dash line respectively represent the simulating results in the condition of 99$\%$ and 90$\%$ transfer efficiency. (f) Simulating cumulative success rate ($R_{c}=\eta^{N}$) of the final atom array as the number of filled sites with 99$\%$ transfer efficiency.}
\label{fig:fig5}
\end{figure*}


All these three sorting algorithms seem feasible and effective to prepare 2D defect-free atom arrays of small size, however their costs of transfer moves and distance are different especially when the atom size is increased. As simulating results, HPFA provides the shortest transfer distance (see Fig.~\ref{fig:fig5}(a)) but costs the most $N_{m}$ (see Fig.~\ref{fig:fig5}(b)). In contrast, we adopt HCA with the fewest $N_{m}$ at the expense of the longest transfer distance since the transfer distance has no obvious relation to the filling fraction of the final atom arrays. HCA will also be a better sorting algorithm for near term intermediate scale atom arrays with 196 atoms compared to HPFA and ASA due to a significantly reduced $N_{m}$ as seen Fig.~\ref{fig:fig5}(c). That the simulating number of moves is almost equal to the experimental result is verified by preparing a $5\times6$ atom array. Additionally, we note that transfer time is also a main factor to affect the filling fraction because of the atom loss in tweezer array limited by the lifetime. For example, if we prepare a $30\times30$ atom array (the details of transfer process has been described above), ASA, HCA and HPFA respectively take 6.7 s, 8.2 s and 9.6 s. This indicates ASA may be the best choice when transfer time dominates the filling fraction. Our time-dependent transfer process is not optimal and we do not take the transfer time into account when selecting a the sorting algorithm, because we have not implemented the fast and adiabatic transport approach yet which can significantly reduces the time cost~\cite{Chen2011,Hickman2020}.

Furthermore, we experimentally measure (see Fig.~\ref{fig:fig5}(d)) and simulate (see Fig.~\ref{fig:fig5}(e)) the filling fractions $\eta$ of defect-free atom arrays using these three sorting algorithms. The $\eta$ is given by $\eta=1-n(1-\zeta)$, where $\zeta$ is the transfer efficiency and $n$ represents the move number per one filled target site. Although the distinction of the filling fraction is small when the transfer efficiency is high and the array size is not large, the sorting algorithm with fewer $n$ demonstrates obvious advantage as the transfer efficiency is decreased or the array size is enlarged. The experimental filling fraction decays as the filled sites number increased mainly induced by the nonuniform transfer efficiency of 64 sites (the efficiency of transporting atoms to the edge sites in atom array is lower than the transfer efficiency to the center sites resulted from imperfect overlap between MT and the edge sites in tweezer arrays). Additionally, we find that the simulating filling fractions with HCA and ASA are approximate to constants as filled site number increased, while HPFA shows a rapidly decline, which means HPFA is not appropriate for large scaled atom arrays even with a high transfer efficiency. Additionally, since the cumulative success rate of the final atom assembler is greatly affected by sorting algorithms as shown Fig.~\ref{fig:fig5}(f), selecting the best one is crucial. To achieve large scale atom assembler with high filling fraction in future, an sufficient laser power is required. The tweezer array generated by our scheme can be scaled to $32\times32$ sites as shown in Fig.~\ref{fig:fig1}(c), but the size of atom assemblers is limited by the optical power of 808 nm laser beam for reliable atom loading.

\section{Conclusion}

In conclusion, to improve the filling fraction of defect-free atom array, we propose a sorting algorithm (HCA) which significantly reduces the sorting moves to near $N$. We experimentally demonstrate the advantage of high filling fraction with HCA over other sorting algorithms and simulate the novel feature of HCA that it can maintain a high and uniform filling fraction as the size of atom assembler increased. This property makes our approach well suited for simulations of quantum dynamics~\cite{Choi2019} and gauge theories~\cite{Celi2020}, error-corrected quantum computations~\cite{Fowler2012,Auger2017}, preparation of single molecule arrays~\cite{Anderegg2019,Zhang2020,He2020} and quantum metrologies based on large scale reconfigurable atom arrays.

\section{acknowledgments}
This work was supported by the National Key Research and Development Program of China under Grant Nos.2017YFA0304501, 2016YFA0302800, and 2016YFA0302002, the National Natural Science Foundation of China under Grant No.11774389, No.12004395 and the Strategic Priority Research Program of the Chinese Academy of Sciences under Grant No.XDB21010100.

C. S. and J. H. contributed equally to this work.

\emph{Note added}.-Recently, we became aware of related work preparing enhanced atom-by-atom assembly~\cite{Schymik2020}.


\begin{thebibliography}{99}

\bibitem{Bernien2017} H. Bernien, S. Schwartz, A. Keesling, H. Levine, A. Omran, H. Pichler, S. Choi, A. S. Zibrov, M. Endres, M. Greiner, V. Vuleti\'{c}, and M. D. Lukin, Probing many-body dynamics on a 51-atom quantum simulator, Nature (London) {\bf 551}, 579 (2017).

\bibitem{Keesling2019}A. Keesling, A. Omran, H. Levine, H. Bernien, H. Pichler, S. Choi, R. Samajdar, S. Schwartz, P. Silvi, S. Sachdev, P. Zoller, M. Endres, M. Greiner, V. Vuleti\'{c}, and M. D. Lukin, Quantum Kibble-Zurek mechanism and critical dynamics on a programmable Rydberg simulator, Nature (London) {\bf 568}, 207 (2019).

\bibitem{deLeseleuc2018} S. de L\'{e}s\'{e}leuc, V. Lienhard, P. Scholl, D. Barredo, S. Weber, N. Lang, H. P. B\"uchler, T. Lahaye, and A. Browaeys, Observation of a symmetry-protected topological phase of interacting bosons with Rydberg atoms, Science {\bf 365}, 775 (2018).

\bibitem{Browaeys2020} A. Browaeys and T. Lahaye, Many-body physics with individually controlled Rydberg atoms, Nat. Phys. {\bf 16}, 132 (2020).

\bibitem{Norcia2019} M. A. Norcia, A. W. Young, W. J. Eckner, E. Oelker, J. Ye, and A. M. Kaufman, Seconds-scale coherence on an optical clock transition in a tweezer array, Science {\bf 366}, 93 (2019).

\bibitem{Madjarov2019} I. S. Madjarov, A. Cooper, A. L. Shaw, J. P. Covey, V. Schkolnik, T. H. Yoon, J. R. Williams, and M. Endres, An Atomic-Array Optical Clock with Single-Atom Readout, Phys. Rev. X 9, 041052 (2019).

\bibitem{Xia2015} T. Xia, M. Lichtman, K. Maller, A. W. Carr, M. J. Piotrowicz, L. Isenhower, and M. Saffman, Randomized benchmarking of single-qubit gates in a 2D array of neutral-atom qubits, Phys. Rev. Lett. {\bf 114}, 100503 (2015).

\bibitem{Wang2017} Y. Wang, A. Kumar, T. Wu, and D. S. Weiss, Single-qubit gates based on targeted phase shifts in a 3D neutral atom array, Science {\bf 352}, 1562 (2017).

\bibitem{Sheng2018} C. Sheng, X. D. He, P. Xu, R. J. Guo, K. P. Wang, Z. Y. Xiong, M. Liu, J. Wang, and M. S. Zhan, HighFidelity Single-Qubit Gates on Neutral Atoms in a TwoDimensional Magic-Intensity Optical Dipole Trap Array, Phys. Rev. Lett. {\bf 121}, 240501 (2018).

\bibitem{Levine2019} H. Levine, A. Keesling, G. Semeghini, A. Omran, T. T. Wang, S. Ebadi, H. Bernien, M. Greiner, V. Vuleti\'{c},H. Pichler, and M. D. Lukin, Parallel Implementation of HighFidelity Multiqubit Gates with Neutral Atoms, Phys. Rev. Lett. {\bf 123}, 170503 (2019).

\bibitem{Graham2019} T. M. Graham, M. Kwon, B. Grinkemeyer, Z. Marra, X. Jiang, M. T. Lichtman, Y. Sun, M. Ebert, and M. Saffman, Rydberg-Mediated Entanglement in a Two-Dimensional Neutral Atom Qubit Array, Phys. Rev. Lett. {\bf 123}, 230501 (2019).

\bibitem{Madjarov2020} I. S. Madjarov, J. P. Covey, A. L. Shaw, J. Choi, A. Kale, A. Cooper, H. Pichler, V. Schkolnik, J. R. Williams and M. Endres. High-fidelity entanglement and detection of alkaline-earth Rydberg atoms. Nat. Phys. {\bf 16}, 857 (2020).

\bibitem{Kim2016} H. Kim, W. Lee, H. G. Lee, H. Jo, Y. Song, and J. Ahn, In situ single-atom array synthesis using dynamic holographic optical tweezers, Nat. Commun. {\bf 7}, 13317 (2016).

\bibitem{Endres2016} M. Endres, H. Bernien, A. Keesling, H. Levine, E. R. Anschuetz, A. Krajenbrink, C. Senko, V. Vuletic, M. Greiner, and M. D. Lukin, Atom-by-atom assembly of defect-free one-dimensional cold atom arrays, Science {\bf 354}, 1024 (2016).

\bibitem{Barredo2016} D. Barredo, S. L\'{e}s\'{e}leuc, V. Lienhard, T. Lahaye, and A. Browaeys, An atom-by-atom assembler of defect-free arbitrary two-dimensional atomic arrays, Science {\bf 354}, 1021 (2016).

\bibitem{Lee2017} W. Lee, H. Kim, and J. Ahn, Defect-Free Atomic Array Formation Using the Hungarian Matching Algorithm, Phys. Rev. A {\bf 95}, 053424 (2017).

\bibitem{Brown2019} M. O. Brown, T. Thiele, C. Kiehl, T.-W. Hsu, and C. A. Regal, Gray-Molasses Optical-Tweezer Loading: Controlling Collisions for Scaling Atom-Array Assembly, Phys. Rev. X {\bf 9}, 011057 (2019).

\bibitem{Mello2019} D. O. de Mello, D. Sch\"affner, J. Werkmann, T. Preuschoff, L. Kohfahl, M. Schlosser, and G. Birkl, Defect-Free Assembly of 2D Clusters of More Than 100 Single-Atom Quantum Systems, Phys. Rev. Lett. {\bf 122}, 203601 (2019).

\bibitem{Yang2016}J. H. Yang, X.D. He, R.J. Guo, P. Xu, K.P. Wang, C. Sheng, M. Liu, J. Wang, A. Derevianko, and M.S. Zhan, Coherence Preservation of a Single Neutral Atom Qubit Transferred between Magic-Intensity Optical Traps, Phys. Rev. Lett. {\bf 117}, 123201 (2016).

\bibitem{Guo2020} R.J. Guo, X.D. He, C. Sheng, J. H. Yang, P. Xu, K.P. Wang, J. Q. Zhong, M. Liu, J. Wang, and M. S. Zhan, Balanced Coherence Times of Atomic Qubits of Different Species in a Dual $3\times3$ Magic-Intensity Optical Dipole Trap Array, Phys. Rev. Lett. {\bf 124}, 153201 (2020).

\bibitem{Li2019} G. Li, Y. L. Tian, W. Wu, S. K. Li, X. Y. Li, Y. X. Liu, P. F. Zhang, and T. C. Zhang, Triply Magic Conditions for Microwave Transition of Optically Trapped Alkali-Metal Atoms, Phys. Rev. Lett. {\bf 123}, 253602 (2019).

\bibitem{Urban2009} E. Urban, T. A. Johnson, T. Henage, L. Isenhower, D. D. Yavuz, T. G. Walker, and M. Saffman, Observation of Rydberg blockade between two atoms, Nat. Phys. {\bf 5}, 110 (2009).

\bibitem{Saffman2010} M. Saffman, T.G. Walker, and K. M\"{o}lmer, Quantum information with Rydberg atoms, Rev. Mod. Phys. {\bf 82}, 2313 (2010).

\bibitem{Wilk2010} T. Wilk, A. Gaetan, C. Evellin, J. Wolters, Y. Miroshnychenko, P. Grangier, and A. Browaeys, Entanglement of Two Individual Neutral Atoms Using Rydberg Blockade, Phys. Rev. Lett. {\bf 104}, 010502 (2010).

\bibitem{Zeng2017} Y. Zeng, P. Xu, X. D. He, Y. Y. Liu, M. Liu, J. Wang, D. J. Papoular, G. V. Shlyapnikov, and M. S. Zhan, Entangling Two Individual Atoms of Different Isotopes via Rydberg Blockade, Phys. Rev. Lett. {\bf 119}, 160502 (2017).

\bibitem{Weiss2017} D. S. Weiss and M. Saffman, Quantum computing with neutral atoms, Phys. Today {\bf 70}, No. 7, 44 (2017).

\bibitem{Saffman2016} M. Saffman, Quantum computing with atomic qubits and Rydberg interactions: progress and challenges, J. Phys. B: At. Mol. Opt. Phys. {\bf49}, 202001 (2016).

\bibitem{Barredo2018} D. Barredo, V. Lienhard, S. de L\'es\'eleuc, T. Lahaye, and A. Browaeys, Synthetic three-dimensional atomic structures assembled atom by atom, Nature (London) {\bf 561}, 79 (2018).

\bibitem{Kumar2018} A. Kumar, T. Wu, F. Giraldo, and D. S. Weiss, Sorting ultracold atoms in a three-dimensional optical lattice in a realization of Maxwell's demon, Nature (London) {\bf 561},83 (2018).

\bibitem{Wang2020} Y. B. Wang, S. Shevate, T. M. Wintermantel, M. Morgado , G. Lochead and S. Whitlock, Preparation of hundreds of microscopic atomic ensembles in optical tweezer arrays, Npj Quantum Inf. {\bf 6}, 1 (2020).

\bibitem{Savard1997} T. A. Savard, K. M. O'Hara, and J. E. Thomas, Laser-noiseinduced heating in far-off resonance optical traps, Phys. Rev. A {\bf 56}, R1095 (1997).

\bibitem{Tuchendler2008} C. Tuchendler, A. M. Lance, A. Browaeys, Y. R. P. Sortais, and P. Grangier, Energy distribution and cooling of a single atom in an optical tweezer, Phys. Rev. A {\bf 78}, 033425 (2008).

\bibitem{Chen2011} X. Chen, E. Torrontegui, D. Stefanatos, J.-S. Li, and J. G.Muga, Optimal trajectories for efficient atomic transport without final excitation, Phys. Rev. A {\bf 84}, 043415 (2011).

\bibitem{Hickman2020} G. T. Hickman and M. Saffman, Speed, retention loss, and motional heating of atoms in an optical conveyor belt, Phys. Rev. A {\bf 101}, 063411 (2020).

\bibitem{Choi2019} S. Choi, C. J. Turner, H. Pichler, W. W. Ho, A. A. Michailidis, Z. Papi\'{c}, M. Serbyn, M. D. Lukin, and D. A. Abanin, Emergent SU(2) Dynamics and Perfect Quantum Many-Body Scars, Phys. Rev. Lett. {\bf 122}, 220603 (2019).

\bibitem{Celi2020} A. Celi, B. Vermersch, O. Viyuela, H. Pichler, M. D. Lukin, and P. Zoller, Emerging Two-Dimensional Gauge Theories in Rydberg Configurable Arrays, Phys. Rev. X {\bf 10}, 021057 (2020).

\bibitem{Fowler2012} A. G. Fowler,J. M. Auger, S. Bergamini and D. E. Browne, Surface codes: Towards practical large-scale quantum computation, Phys. Rev. A {\bf 86}, 032324 (2012).

\bibitem{Auger2017} J. M. Auger, S. Bergamini, and D. E. Browne, Blueprint for fault-tolerant quantum computation with Rydberg atoms, Phys. Rev. A {\bf 96}, 052320 (2017).

\bibitem{Anderegg2019} L. Anderegg, L. W. Cheuk, Y. C. Bao, S. Burchesky, W. Ketterle, K.-K. Ni, and J. M. Doyle, An optical tweezer array of ultracold molecules, Science {\bf 365}, 1156 (2019).

\bibitem{Zhang2020} J. T. Zhang, Y. Yu, W. B. Cairncross, K. Wang, L. R. B. Picard, J. D. Hood, Y.-W. Lin, J. M. Hutson, and K.-K. Ni, Forming a Single Molecule by Magnetoassociation in an Optical Tweezer, Phys. Rev. Lett. {\bf 124}, 253401 (2020).

\bibitem{He2020} X. D. He, K. P. Wang, J. Zhuang, P. Xu, X. Gao, R. J. Guo, C. Sheng, M. Liu, J. Wang, J. M. Li, G. V. Shlyapnikov, M. S. Zhan, Coherently forming a single molecule in an optical trap, Science {\bf 370}, 331 (2020).

\bibitem{Schymik2020} K. N. Schymik, V. Lienhard, D. Barredo, P. Scholl, H. Williams, A. Browaeys and T. Lahaye, Enhanced atom-by-atom assembly of arbitrary tweezers arrays, arXiv:2011.06827v1.
\end{thebibliography}

\end{document}